\documentclass[aps,prl,twocolumn,showpacs,amsmath,amssymb,floatfix,superscriptaddress,notitlepage]{revtex4-2}
\usepackage[utf8]{inputenc}
\usepackage{graphicx}
\usepackage{xcolor}

\usepackage{bm}
\begin{document}

\author{Nathan Shettell}
%\email{}
\affiliation{Centre for Quantum Technologies, National University of Singapore, Singapore}

\author{Rainer Dumke}
\affiliation{Centre for Quantum Technologies, National University of Singapore, Singapore}
\affiliation{School  of  Physical  and  Mathematical  Sciences,  Nanyang  Technological  University,  637371, Singapore}

%\flushbottom
\title{Emulating an Atomic Gyroscope with Multiple Accelerometers}

\begin{abstract}
The main advantage of an atomic accelerometer when compared to a classical accelerometer is negligible bias drift, allowing for stable long-term measurements, which opens the potential application in navigation. This negligible drift arises from the fact that the measurements can be traced back to natural constants, and the system is intrinsically stable due to the simple design. In this manuscript, we extend this property of long-term stability to gyroscopic measurements by considering an array of atomic accelerometers, and comparing the performance to atomic gyroscopes, which are technologically more prone to bias drifts. We demonstrate that an array consisting of four three-axis atomic accelerometers can outperform state of the art atomic gyroscopes with respect to long-term stability.
\end{abstract}

\maketitle

\section*{Introduction}

Atomic interferometry has undergone rapid technological developments in the past few decades \cite{kasevich1991atomic, baudon1999atomic, cronin2009optics, kitching2011atomic}, leading to highly sophisticated technologies that can be used for a variety of precision measurements, including, but not limited to, Earth's gravity \cite{peters2001high, bidel2020absolute, oon2022compact, bertoldi2006atom, sorrentino2014sensitivity}, the fine structure constant \cite{parker2018measurement, morel2020determination} and the Aharonov–Casher effect \cite{sangster1993measurement}. Another promising application of atomic intereferometers is navigation and inertial sensing \cite{canuel2006six, battelier2016development, lautier2014hybridizing, fang2012advances, jiang2018parametrically, feng2019review, geiger2020high, templier2022tracking,tennstedt2021integration,tennstedt2023atom}, i.e. as an accelerometer or gyroscope. The main justification for atomic navigation devices (as opposed to a classical device) is that the precision of the measurement can rival the best (classical) inertial sensors \cite{fang2012advances, tazartes2014historical, el2020inertial, narducci2022advances} while maintaining accuracy for a much longer period of time.

Bias drifts are a common hindrance to classical sensors, mainly due to temperature dependencies within the hardware, mechanical wear-and-tear, and aging, which ultimately degrade the accuracy of long-term measurements. In comparison, the measurements taken with atomic accelerometers are based on very well-known and controlled scale factors: the inter-pulse duration, laser wavelength and transition energies, which allows, in principle, for long-term absolute measurements \cite{niebauer1995new, bidel2020absolute, oon2022compact, lautier2014hybridizing, yankelev2019multiport}. Despite atomic gyroscopes depending on the same aforementioned factors, the mechanical and optical setup is more complex, consequently, the atom trajectory drifts over time, limiting the long-term stability \cite{fils2005influence, gauguet2008off, gauguet2009characterization}. The bias drifts exhibit in gyroscopes vary over several orders magnitude, from $~10^{-5}-10^{-9}\;$rad$\cdot$s$^{-1}$/day \cite{durfee2006long, canuel2006six, gustavson2000rotation, jiang2018parametrically,  stockton2011absolute, gauguet2008off, berg2015composite, yao2018calibration, dutta2016continuous, savoie2018interleaved}. A stability limit of $\sim 7 \times 10^{-8}\;$rad$\cdot$s$^{-1}$/day was hypothesized for atomic gyroscopes, due to wavefront aberrations and off-resonant Raman transitions \cite{fils2005influence, gauguet2008off}. Notably though, this limit has been surpasses by the gyroscopes demonstrated in Refs.~\cite{dutta2016continuous, savoie2018interleaved}.

In this work, we explore the prospect of measuring rotational motion with sufficiently many atomic accelerometers \cite{padgaonkar1975measurement, madgwick2013measuring, pachter2013gyro, nilsson2016inertial}. Henceforth, such a contraption is referred to as an array of atomic accelerometers (AAA). To compute all three vector components of the angular velocity, one must utilize both vertical and horizontal atomic accelerometers. Although vertical atomic accelerometers do exhibit negligible drift (for example, no drifting properties were witnessed after 24 hours of measurements in Ref.~\cite{oon2022compact}) the same is not true for horizontal accelerometers \cite{canuel2006six, geiger2011detecting, xu2017quantum, perrin2019zero, templier2022tracking}. From a technological viewpoint, horizontal and vertical accelerometers are effectively equivalent. Thus one can imagine that as horizontal accelerometers garner more research attention, they could similarly exhibit negligible bias drifts. In the meantime, this article explores how a bias drift present in a three-axis accelerometer propagates to the gyroscopic measurements of an AAA.

The working principle of an AAA is to deduce rotational velocities from linear acceleration gradients. This has been implemented with great success for navigation \cite{tan2001design, zorn2002gps, buhmann2006gps, qin2006attitude, naseri2014improving} because of the high bandwidth of classical accelerometers. Notably, some of the navigational devices are GPS-aided \cite{zorn2002gps, buhmann2006gps}; long-term stability offered by an AAA is vital to allow for navigation in GPS denied environments. Other interesting applications of arrays of accelerometers include bio-mechanics \cite{nusholtz1984head, beckwith2007validation}, stabilization \cite{algrain1993accelerometer, latt2011placement, lee2003increasing} and seismology \cite{niazi1986inferred, oliveira1989rotational, spudich1995transient, huang2003ground}, the latter of which can be is a potential application for an AAA. Evidently, the long-term stability of atomic accelerometers is a favourable quality when measuring dynamics on longer timescales imposed by nature, for example, volcanic activity, and underground geothermal activity and glacier ablation.

It is worthwhile to note that using multiple atomic accelerometers does not multiplicatively increase the technological overhead, for example, the same laser system and control electronics can be used for multiple sensor heads. Such a contraption would provide a complete navigational description (linear and rotational motion) using a sole type of technology, facilitating maintenance and troubleshooting. Atomic accelerometers vary drastically in size and performance, but the working principle is the same: an atomic cloud with discrete energy states is stimulated by a sequence of laser pulses, the phase shift accumulated through two different paths of the interferometer is proportional to the projection of the acceleration of the atom relative to an inertial reference along the interrogation laser direction \cite{kitching2011atomic}; said phase shift is mapped onto a probability distribution and can be deduced through statistical modelling. A thorough explanation of the underlying physics can be found in Ref. \cite{peters2001high}. As we are investigating this idea from a theoretical perspective, the atomic accelerometers are considered to be a blackbox which outputs an estimate of the linear acceleration with statistical constraints (precision, sensitivity, resolution, stability, et cetera).

This work is organized as follows. We begin by outlining the pertinent mathematics to an array of atomic accelerometers. Next, we compare the performance of an AAA to current atomic gyroscopes \cite{durfee2006long, berg2015composite, gauguet2009characterization, yao2018calibration, savoie2018interleaved}. This is done by considering accelerometers with realistic resolutions \cite{wu2017multiaxis, chen2019single, templier2022tracking} and subjecting the array to different degrees of rotational motion. We investigate how a bias drift within the atomic accelerometers translates to an effective bias drift with respect to the angular velocity measurement. Additionally, we incorporate expected technical challenges, such as size limitations, positional uncertainties, and misalignment, to determine their effects on the attainable resolution.

\section*{Array of accelerometers}

Within a rigid body, the difference of linear acceleration (in the rotating reference frame) observed at locations $\bm{r}^{(1)}$ and $\bm{r}^{(2)}$ is given by
\begin{equation}
    \label{eq:gen_motion}
    \bm{a}^{(1)}-\bm{a}^{(2)}=W(\bm{r}^{(1)}-\bm{r}^{(2)}),
\end{equation}
where
\begin{equation}
    W=\dot{\Omega}+\Omega^2,
\end{equation}
with
\begin{equation}
    \Omega=
    \begin{pmatrix}
        0 && \omega_z && -\omega_y \\
        -\omega_z && 0 && \omega_x \\
        \omega_y && -\omega_x && 0 \\
    \end{pmatrix},
\end{equation}
and $\bm{\omega}=(\omega_x,\omega_y,\omega_z)$ is the angular velocity of the rigid body and $\dot{\Omega}=\frac{\partial \Omega}{\partial t}$. It follows that, with sufficiently many accelerometers one can measure the linear acceleration at different locations on the rigid body and reverse engineer said measurements to determine the angular velocity $\bm{\omega}$ \cite{padgaonkar1975measurement, schuler1967measuring, madgwick2013measuring, pachter2013gyro, nilsson2016inertial, carlsson2022inertial}.

\begin{figure}
    \centering
    \includegraphics[width=0.44\textwidth]{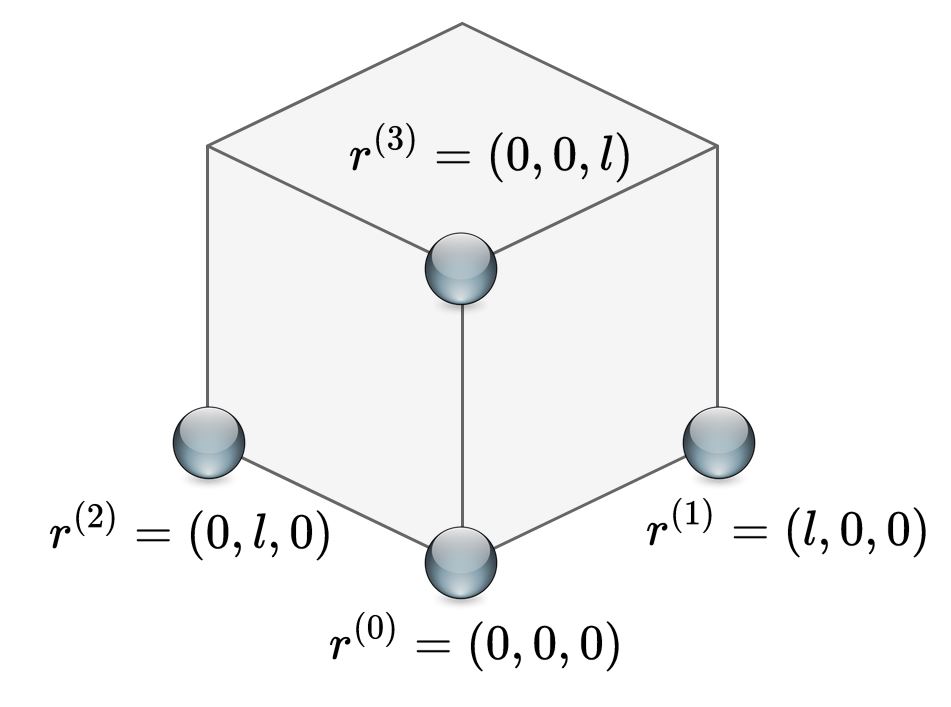}
\caption{This figure illustrates the AAA we focus on in this study: the location of one central accelerometer is denoted the origin and the other three are each placed along a separate axis at a distance $l$ from the origin. We assume that the atomic accelerometers are simultaneously sensitive to three different axis of acceleration. Although such devices exist, the measurements are sequential and not simultaneous \cite{canuel2006six, stern2009light, geiger2011detecting, templier2022tracking, barrett2019multidimensional}, however, assuming the time delay between measurements is small, than this approximation may still hold. Alternatively, one can use a different setup with single-axis atomic accelerometers placed strategically \cite{schuler1967measuring} and obtain simultaneous measurement results.}
\label{fig:AccArrayFig}
\end{figure}

A minimum of nine single-axis accelerometers are needed to infer the symmetric terms of $W$ (angular velocity) as well as the anti-symmetric terms (angular acceleration). Because of the quadratic nature of $\Omega^2$, any estimate of  $\bm{\omega}$ has intrinsic sign ambiguity, however this can be eliminated using dependencies between $\bm{\omega}$ and $\dot{\bm{\omega}}$ \cite{williams2013minimal}. In this manuscript we focus on an array of four three-axis atomic accelerometers: where the placement of one of the accelerometers is deemed the origin, and each one of the remaining accelerometers is fixed at a distance $l$ along a separate axis, as depicted in Figure~\ref{fig:AccArrayFig}. This choice is made for mathematical simplicity, nevertheless, much of the analysis can be straightforwardly adopted to arrays of nine single-axis accelerometers \cite{schuler1967measuring}, or of arbitrarily large arrays.

We use the notation $\hat{\square}$ to denote an estimated quantity, e.g. $\hat{W}$ is the estimate of $W$. There are numerous methods of formulating the estimate $\hat{\bm{\omega}}$ from $\hat{W}$ \cite{cardou2008angular, cardou2010nonlinear}, most of them diverge when one or more components of $\bm{\omega}$ approaches zero \cite{parsa2005design}. The methodology we adopt is proposed in Ref.~\cite{cardou2008angular}, where one considers the diagonal terms of $W$ in a rotated reference frame, specifically a reference frame where each component of $\bm{\omega}$ has equal magnitude. This method attains a high degree of precision and diverges solely when $\bm{\omega}$ tends to exactly zero  \cite{cardou2008angular}. An a priori estimate $\tilde{\bm{\omega}}$ is required to determine the correct reference frame, this can be achieved using a classical sensor, or alternatively, one can use the angular acceleration measurement to construct a self-contained estimation strategy
\begin{equation}
    \label{eq:selfcontained}
    \tilde{\bm{\omega}}(t_{n}) = \hat{\bm{\omega}}(t_{n-1})+\frac{\Delta t}{2} \big( \hat{\dot{\bm{\omega}}}(t_{n-1})+\hat{\dot{\bm{\omega}}}(t_n) \big),
\end{equation}
where $t_n$ is used to denote an estimate of a quantity at the $n$th measurement and $\Delta t$ is the time between measurements.

Because of the non-linearity of the estimation, it is difficult to determine exact statistical tendencies. Regardless, a guideline for the error of an array of accelerometers with $N$ three-axis atomic accelerometers is  
\begin{equation}
    \centering
    \label{eq:omega_resolution}
    || \hat{\bm{\omega}}-\bm{\omega}|| \propto \frac{|| \hat{\bm{a}}-\bm{a}||}{\sqrt{M} l_\text{eff} ||\bm{\omega}||},
\end{equation}
where $M$ is the number of ways to select four accelerometers which are non-coplanar, and $l_\text{eff}$ is the average separation distance between pertinent accelerometers. The quantity $M$ is dependent on the arrangement of the accelerometer within the AAA, in addition, other scaling factors may be introduced depending on the employed inference strategy. More information on the statistics of accelerometer arrays can be found in Refs.~\cite{skog2016inertial, madgwick2013measuring, klein2015analytic, cucci2016analysis}.

Evidently, the capabilities of deducing rotational motion will be bounded by the effectiveness of the accelerometers, hence the proportionality to $|| \hat{\bm{a}}-\bm{a}||$, and similarly improves with the number of accelerometers within the array ($N$). Moreover, the precision of the apparatus increases when the acceleration on different sensors vary substantially, which is achieved for large values of $||\bm{\omega}||$, or can be artificially achieved by increasing the distance between sensors ($l$).

\section*{Long-term stability of $\hat{\bm{\omega}}$}

In order to properly compare the long-term stability of the angular velocity measurement made by an AAA, we model the statistical properties of the three-axis atomic accelerometers from Ref.~\cite{templier2022tracking}. Namely a single-shot measurement resolution of $7.6 \times 10^{-5}\;$ms$^{-2}$, and a bias drift of $5.9 \times 10^{-7}\;$ms$^{-2}$/day. Here, the single-shot measurement resolution concerns the norm of the acceleration vector. For the sake of simplicity, we model the resolution of the each vector component (i.e. $\hat{a}_x^{(i)}$, $\hat{a}_y^{(i)}$, and $\hat{a}_z^{(i)}$) to be identical, thus equal to $7.6/\sqrt{3} \times 10^{-5}\;$ms$^{-2}$, which is approximately observed in Ref.~\cite{templier2022tracking}. In addition, we assume the sensitivity of the accelerometers are independent from $\bm{a}$ and $\bm{\omega}$, this is in to more easily isolate properties of the effective bias drift with respect to $\hat{\bm{\omega}}$. If one desired to obtain the bias-drift for a different sensitivity, one could scale the results according to Eq.~\eqref{eq:omega_resolution}.

The data is modeled using
\begin{equation}
    \hat{\bm{a}}^{(i)}=\bm{a}^{(i)}+\bm{\eta}^{(i)}+\beta^{(i)} \bm{u}^{(i)},
\end{equation}
where $\bm{\eta}^{(i)}$ is a random variable to emulate white noise, $\beta^{(i)}$ is the bias of the $i$th sensor and $\bm{u}^{(i)}$ is a unit vector signifying the direction in which the accelerometer drifts towards. To simulate the dynamical properties of the drift, the bias at the $n$th time step, $\beta^{(i)}(t_n)$, is sampled from a normal distribution centered around $\beta^{(i)}(t_{n-1})+\dot{\beta} \Delta t$, where $\dot{\beta}=5.9 \times 10^{-7}$ ms$^{-2}$/day is the drift rate.

For (most) classical inertial sensors, bias drifts are gauged by performing long-term static measurements; thus, the bias drifts are modeled to be independent of the degree of motion. This is not true for an AAA because of the non-linearity present in the method in which $\hat{\omega}$ is constructed. There is an inverse dependency on the angular velocity, $||\bm{\omega}||$. Hence, the simulation is repeated for different values of $\omega$. For the sake of symmetry, the AAA is subjected to constant motion $\bm{\omega}=\omega (1,1,1)/\sqrt{3}$, in which $\omega \in \{ 0.05,0.50,5.00 \}\;$rad$\cdot$s$^{-1}$. Data is sampled for a total time of $T=2 \times 10^{6}\;$s, with a sampling time of $\Delta t = 0.25\;$s. This value of $\Delta t$ is comparable to the interrogation of time of most modern atomic accelerometers \cite{geiger2020high}, and would provide data readout at an acceptable rate for a variety of applications, whether that be navigation or seismology. Typically, the time difference between data readout is a few seconds (e.g. $\Delta t \approx 1.65\;$s in Ref.~\cite{templier2022tracking}) because of dead time, however, techniques have been developed which effectively eliminate dead times \cite{dutta2016continuous}. Note though, that the value of $\Delta t$ has very little impact on the performance of the AAA. The exception of course is the construction of the a priori $\tilde{\bm{\omega}}$ in the self-contained strategy, Eq.~\eqref{eq:selfcontained}, however, the error added by a marginally worse choice of $\tilde{\bm{\omega}}$ is greatly overshadowed by white noise; in the event that Eq.~\eqref{eq:selfcontained} is problematic, one could instead use a classical sensor as a substitute for~$\tilde{\bm{\omega}}$.

\begin{figure*}
    \centering
    \begin{minipage}{0.495\textwidth}
        \centering
        \includegraphics[width=\textwidth]{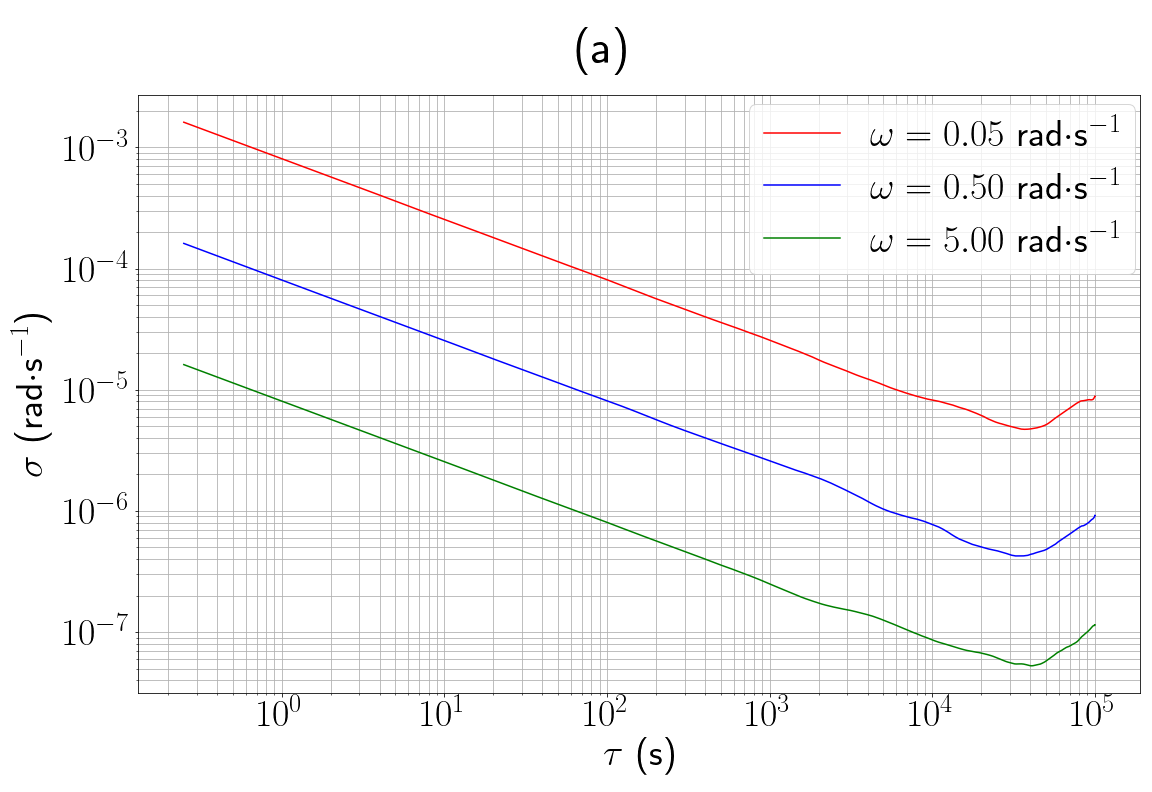}
    \end{minipage}\hfill
    \begin{minipage}{0.495\textwidth}
        \centering
        \includegraphics[width=\textwidth]{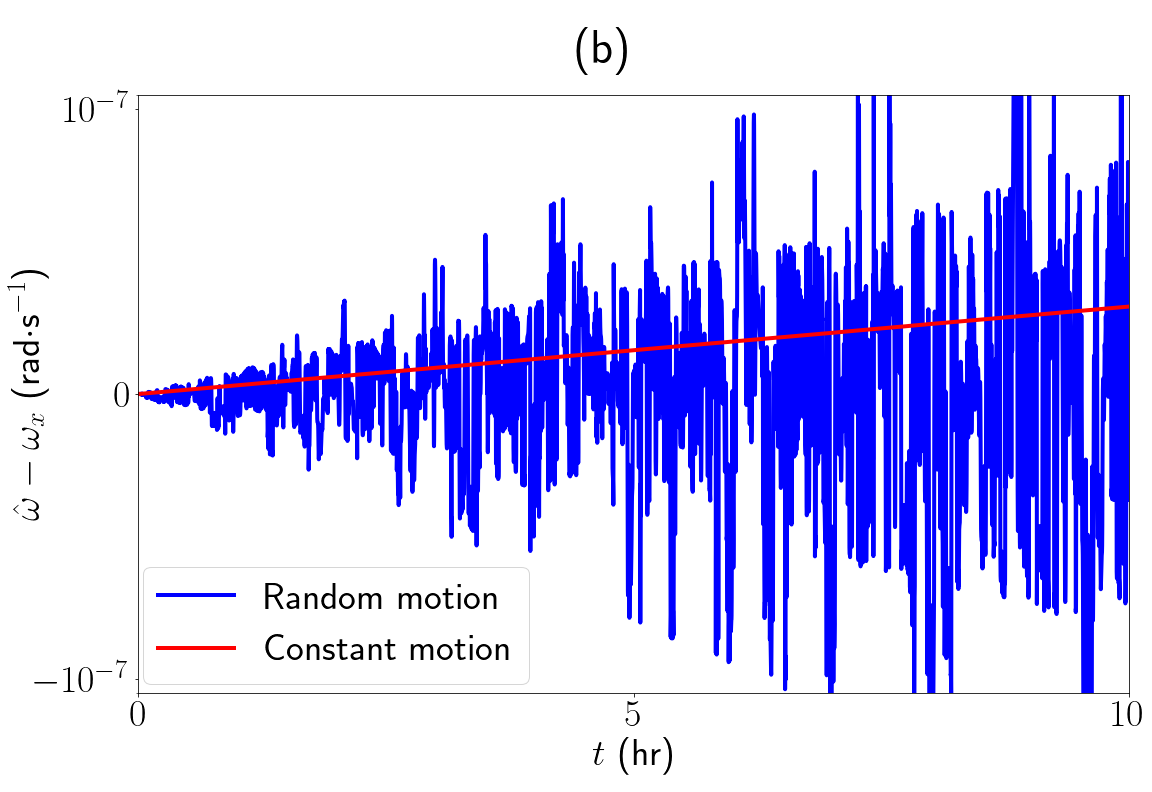}
    \end{minipage}
    \caption{(a) Long-term stability analysis for the angular velocity measurement, in which the accelerometers are subjected to a bias with a drift rate of $6 \times 10^{-7}$ ms$^{-2}$/day. The accelerometers are subjected to a constant rotation of $\bm{\omega}=\omega (1,1,1)/\sqrt{3}$ and are separated by a distance of $l=1\;$m. The total Allan deviation $\sigma = (\sigma_x^2+\sigma_y^2+\sigma_z^2)^{1/2}$ is plotted, where $\sigma_x,\sigma_y,\sigma_z$ are the Allan deviations corresponding to their respective vector component of $\hat{\bm{\omega}}$. (b) Error analysis of $\hat{\omega}_x$, for $l=1\;$m and $||\bm{\omega}||=1\;$rad$\cdot$s$^{-1}$. The white noise terms, $\bm{\eta}^{(i)}$, are omitted to highlight the effects of the bias drift. When the array undergoes constant rotation, the drift behaves similarly to a traditional inertial sensor, however, when the array undergoes random motion, the effective drift behaves as white noise which increases in severity with time.}
    \label{fig:DriftAnalysis}
\end{figure*}

As a consequence of the subtraction in Eq.~\eqref{eq:gen_motion}, the severity of the bias ultimately depends $\bm{u}^{(i)}-\bm{u}^{(j)}$. Hence, if two accelerometers drift in the same direction, the quantity $\hat{\bm{a}}^{(i)}-\hat{\bm{a}}^{(j)}$ will not induce any drift on the gyroscopic measurements. On the other hand, the effective bias drift will be largest when $\bm{u}^{(i)}=-\bm{u}^{(j)}$. For the particular layout we are simulating, see Figure \ref{fig:AccArrayFig}, the worst case scenario occurs when the bias of each axial sensor is opposite to the central sensor, which is what we consider in the simulation, Figure~\ref{fig:DriftAnalysis}a.

%Regardless of the values of $\omega$ and $l$, we find that the bias becomes the limiting factor in accuracy after approximately $T \approx 3 \times 10^4\;$s of interrogation. For $\omega=5\;$ rad$\cdot$s$^{-1}$, the effective bias drift exhibited in the angular velocity measurements is $4.3 \times 10^{-8}\;$ rad$\cdot$s$^{-1}$/day when $l=2\;$m and $1.2 \times 10^{-7}\;$ rad$\cdot$s$^{-1}$/day when $l=1\;$m. For $\omega=1\;$ rad$\cdot$s$^{-1}$, the effective bias drift exhibited in the angular velocity measurements is $2.9 \times 10^{-7}\;$ rad$\cdot$s$^{-1}$/day when $l=2\;$m and $4.3 \times 10^{-7}\;$ rad$\cdot$s$^{-1}$/day when $l=1\;$m. The first three values listed surpass the stability limits of the majority of modern atomic gyroscopes \cite{berg2015composite, gauguet2009characterization, yao2018calibration, durfee2006long}. Furthermore, this simulation assumes the worst case scenario, in which  the bias drifts of the radial sensors occur in the exact opposite direction to the central sensor; on average one expects an improvement by a factor of $\sqrt{2}$.

Regardless of the values of $\omega$, we find that the bias becomes the limiting factor in accuracy after approximately $T \approx 3 \times 10^4\;$s of interrogation. The efficacy of Eq.~\eqref{eq:omega_resolution} is showcased in the simulation results of Figure~\ref{fig:DriftAnalysis}a, as increasing $\bm{\omega}$ by a factor of ten, similarly improves the effective bias drift by a factor of ten. The most-promising result concerning the simulation when $\omega=5.00\;$ rad$\cdot$s$^{-1}$, in which the corresponding bias drift exhibited is approximately $1.2 \times 10^{-7}\;$ rad$\cdot$s$^{-1}$/day. This value surpasses the stability limits of the majority of modern atomic gyroscopes \cite{berg2015composite, gauguet2009characterization, yao2018calibration, durfee2006long} and is suitable for the navigation of maritime vehicles \cite{el2020inertial}. For the chosen configuration of atomic accelerometers with $l=1\;$m, the long-term stability of the AAA is not suitable for analyzing rotational processes with small $|\bm{\omega}|$ over a long period of time, such as seismology. Nevertheless, the effective long-term stability of the AAA can be enhanced by increasing $l$ or by introducing additional sensors. Moreover, the simulation performed assumed the worst case scenario, in which  the bias drifts of the radial sensors occur in the exact opposite direction to the central sensor; on average one expects an improvement by a factor of $\sqrt{2}$.

Evidently, when $||\bm{\omega}||$ is increased, then the effective bias drift is reduced. One technique which could ensure this to be the case, is to subject the AAA to a fixed rotation with angular velocity $\bm{\omega}_0$ such that $||\bm{\omega}_0||^2 \gg ||\hat{\bm{a}}^{(i)}-\bm{a}^{(i)}||/l$ for all $i$. If the unknown angular velocity is negligible in comparison, i.e. $||\bm{\omega}_0+ \bm{\omega}|| \approx ||\bm{\omega}_0||$, then $||\bm{\omega}_0||$ will dictate the bias drift present within the measurements. Evidently, one must be able to correctly set $\bm{\omega}_0$ with a a higher degree of precision compared to the desired precision of $\hat{\omega}$, otherwise the sensitivity of $\hat{\omega}$ will retrogress. Additionally, one must be mindful of the dynamic range of the accelerometers \cite{narducci2022advances}, otherwise, the atomic accelerometers may become unusable due to complete loss of contrast \cite{lan2012influence}. Hence, the choice of $\bm{\omega}_0$ will greatly dependent on the technological specifications of the hardware and intended use case of the AAA.

Notably, drifting accelerometers has a vastly different effect on $\hat{\bm{\omega}}$ when the motion is not constant, as portrayed in Figure~\ref{fig:DriftAnalysis}b. When $\bm{\omega}$ is kept fixed, the effective drift behaves as expected, however, for random motion (see Figure~\ref{fig:OmegaValues}) the effective drift behaves as white noise which becomes progressively more severe as time increases.

\section*{Resolution analysis}

\begin{figure*}
    \centering
    \includegraphics[width=0.75\textwidth]{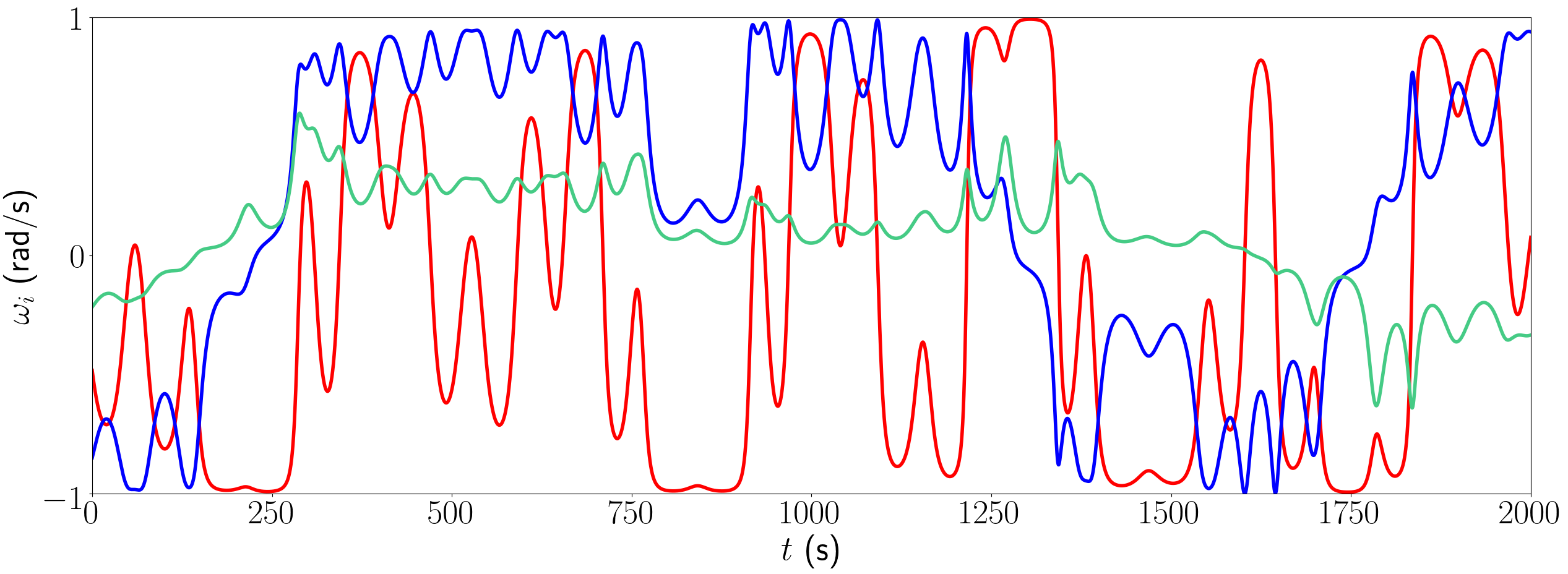}
\caption{The simulated random values used for $\omega_x$ (red), $\omega_y$ (blue) and $\omega_z$ (green), scaled such that $||\bm{\omega}||=1$.}
\label{fig:OmegaValues}
\end{figure*}

In this section we gauge how different technical challenges inherit to an AAA influence its resolution, namely size limitations, positional uncertainties of the individual sensors, and misalignment of the sensor heads. In all of the simulations, the AAA is subjected to random motion, shown in Figure~\ref{fig:OmegaValues}, where $||\bm{\omega}||$ is kept fixed.

When simulating the effects of positional uncertainties and the misalignment, we are considering these to be unknown errors which are static in time. In reality, these values may drift over time leading to another source of bias drift and ultimately effecting the long-term stability. However, if one periodically measures these values, then any change can be properly accounted for when calculating $\bm{\omega}$ (up to the precision of the measurement device, which is what we deem as positional uncertainty and misalignment).

\begin{figure*}
    \centering
    \begin{minipage}{0.333\textwidth}
        \centering
        \includegraphics[width=\textwidth]{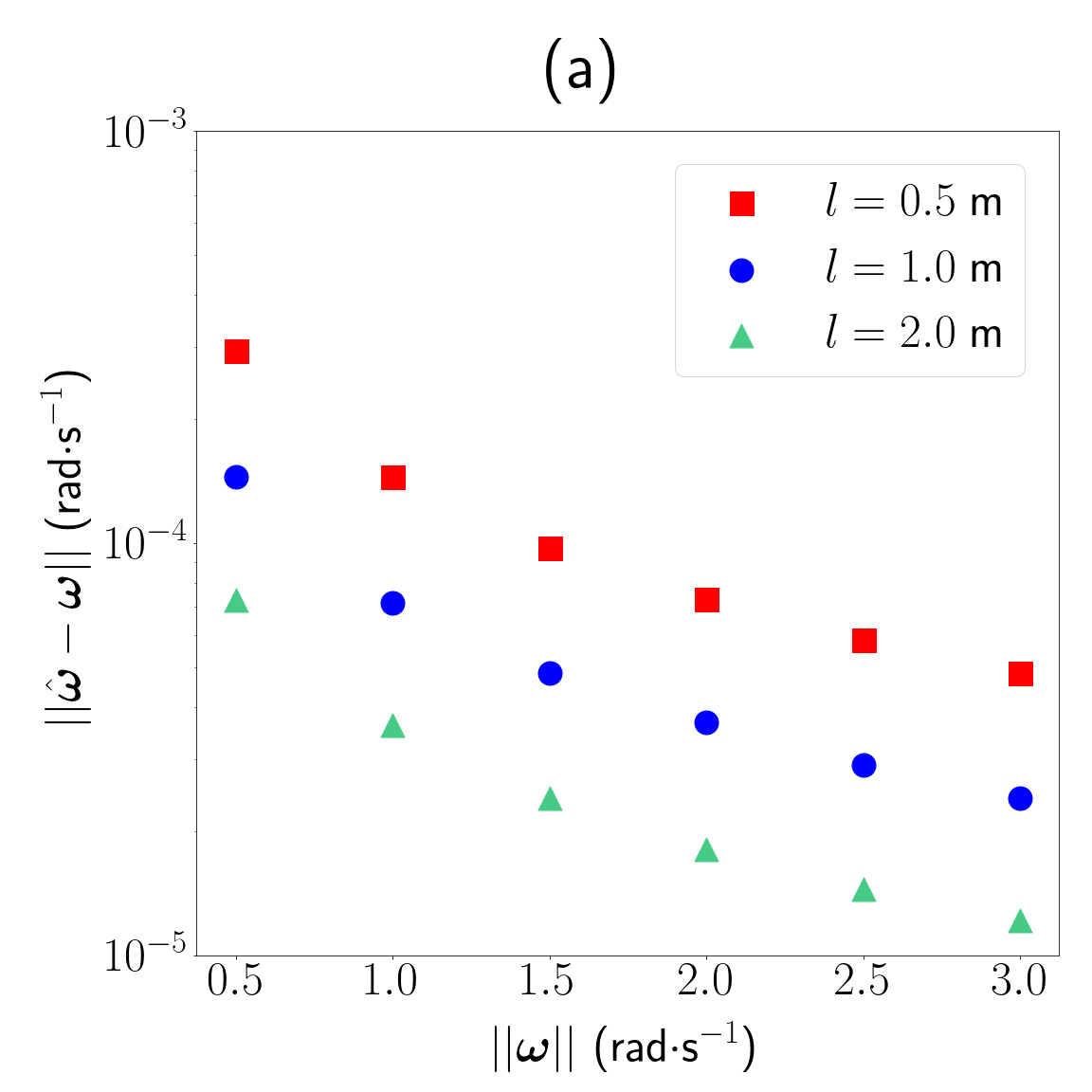}
    \end{minipage}\hfill
    \begin{minipage}{0.333\textwidth}
        \centering
        \includegraphics[width=\textwidth]{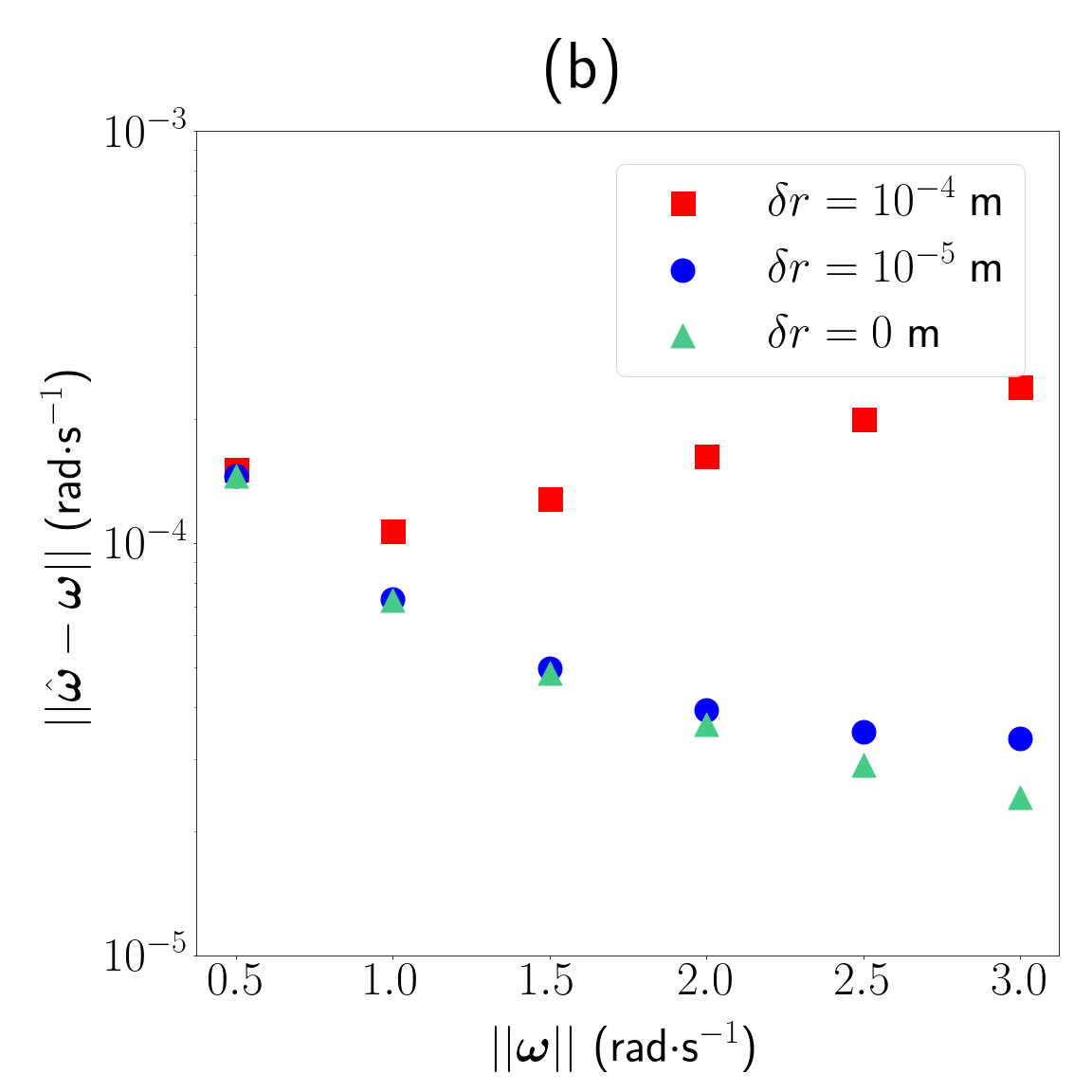}
    \end{minipage}\hfill
    \begin{minipage}{0.333\textwidth}
        \centering
        \includegraphics[width=\textwidth]{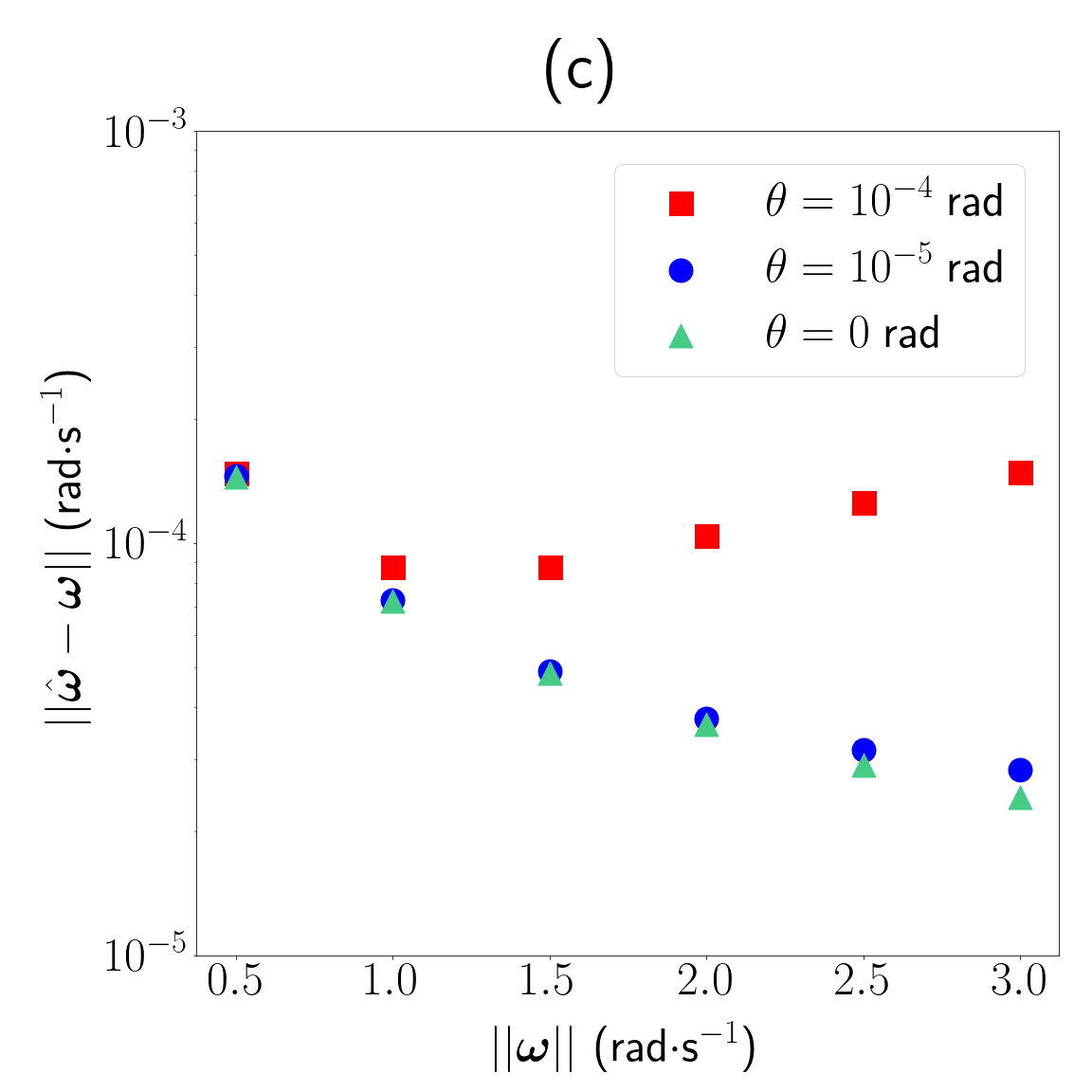}
    \end{minipage}
    \caption{Resolution analysis with respect to different technological challenges inherit to an AAA. In all three plots, the values are obtained by averaging over 25 simulations, this is to consider different values of $\bm{\eta}$, and different directions for the positional uncertainty in (b) and axis of rotation in (c). In each of the 25 simulations, the average error  $||\hat{\bm{\omega}}-\bm{\omega}||$ is computed, where the total interrogation time is $T=2 \times 10^3\;$s and $\Delta t=0.25\;$s. Note that, the error fluctuates minimally in time due to the method in which $\hat{\bm{\omega}}$ is constructed. (a) Resolution analysis for different sizes of an AAA. (b) Resolution analysis for different degrees of positional uncertainty (with $l=1.0\;$m). (c) Resolution analysis for different degrees of misalignment (with $l=1.0\;$m). The trends in (b) and (c) are similar because a positional uncertainty will induce (with high probability) an angular misalignment; the semblance in numerical values is due to the choice $l=1.0\;$m.}
    \label{fig:Resolution}
\end{figure*}

\subsection*{Size Limitations}

The first aspect we explore is how the size of the AAA affects the attainable resolution. Here, the data is simply simulated using
\begin{equation}
    \label{eq:size_sim}
    \hat{\bm{a}}^{(i)}=\bm{a}^{(i)}+\bm{\eta}^{(i)},
\end{equation}
in which different values of $l$ is considered.

Figure \ref{fig:Resolution}a clearly exhibits the hypothesized tendencies for the resolution of the array of atomic accelerometers, Eq.~\eqref{eq:omega_resolution}, in which the error $||\hat{\bm{\omega}}-\bm{\omega}||$ is computed for a variety of values for $l$ and $||\bm{\omega}||$. Even for very large arrays with $l=2.0\;$m, the attainable single-shot resolution of $\sim 10^{-5}\;$rad$\cdot$s$^{-1}$ is greatly outperformed by state of the atomic gyroscopes, whose single-shot resolution is on the order of $\sim 10^{-8} - 10^{-9}\;$rad$\cdot$s$^{-1}$ \cite{savoie2018interleaved}. The single-shot resolution of the AAA can be improved upon by introducing additional accelerometers, however, the most practical way of rivaling atomic accelerometers is by hybridizing the array of accelerometers with classical sensors \cite{lautier2014hybridizing, cheiney2018navigation, geiger2011detecting, merlet2009operating}; note that this does not defeat the purpose of an AAA, as their most impactful contribution is long-term stability.

\subsection*{Positional Uncertainty}

The second aspect we explore is how positional uncertainties in the placement of the individual atomic accelerometers affects the attainable resolution. This positional uncertainty could be caused by fluctuations in the locations of the atomic cloud within the atomic accelerometers. Here, the data is again simulated using Eq.~\eqref{eq:size_sim}, with the addition that the position of the atomic accelerometers, $\bm{r}^{(i)}$, are shifted by a distance $\delta r$ in the direction of a random unit vector $\bm{u}^{(i)}$, hence $\bm{r}^{(i)} \rightarrow \bm{r}^{(i)} + \delta r \bm{u}^{(i)}$.

We can deduce from Figure \ref{fig:Resolution}b that there is some threshold value for $\delta r$, which when surpassed causes the resolution to deviate from the form proposed in Eq.~\eqref{eq:omega_resolution}. For $l=1.0\;$m, this threshold is between $\delta r =10^{-4}\;$m and $\delta r =10^{-5}\;$m. Moreover, the deviation is exacerbated for larger values of $||\bm{\omega}||$. This threshold can be overcome by utilizing high-precision calibration techniques, such as laser interferometry, which can be used to measure distances on a $10^{-5}$\;m scale.

\subsection*{Misalignment}

The third aspect we explore is how misalignment of the individual atomic accelerometers affects the attainable resolution. The misalignment could be caused by fluctuations in the angle of the retro-reflective mirrors present in the atomic accelerometers, or simply mechanical error when positioning each atomic accelerometer. Here, the data is simulated using
\begin{equation}
    \hat{\bm{a}}^{(i)}=M_\theta^{(i)}(\bm{a}^{(i)}+\bm{\eta}^{(i)}),
\end{equation}
where $M_\theta^{(i)}$ is a matrix which induces a rotation of an angle $\theta$ about a random axis.

Similar to the positional uncertainty, $\delta r$, we can deduce from Figure~\ref{fig:Resolution}c that there is some threshold value for $\theta$, which when surpassed causes the resolution to deviate from the form proposed in Eq.~\eqref{eq:omega_resolution}. For $l=1.0\;$m, this threshold is between $\theta =10^{-4}\;$rad and $\theta =10^{-5}\;$rad. Fortunately, mechanical alignment can be measured with extremely high-precision, and similarly retro-reflective mirrors are calibrated on the order of $\sim 10^{-6}\;$rad \cite{peters2001high} and can be stabilized with extremely precise tilt calibration \cite{oon2022vibration}. Hence, the cumulative misalignment for each atomic accelerometer is not expected to surpass the aforementioned threshold.

\section*{Discussion}

The long-term stability of atomic accelerometers has greatly improved in the past few decades \cite{feng2019review, geiger2020high} and currently outclass classical accelerometers. Even though atomic gyroscopes have similarly improved, their long-term stability has not yet surpassed that of classical gyroscopes \cite{feng2019review, fang2012advances, geiger2020high}. In this manuscript, we have demonstrated that an array consisting of four three-axis atomic accelerometers can be an effective substitute for an atomic gyroscope whose long-term stability surpasses the hypothesized stability limit imposed on atomic gyroscopes \cite{fils2005influence, gauguet2008off}. The long-term stability was simulated using parameters describing a recently developed three-axis gyroscope \cite{templier2022tracking}, which exhibits a non-negligible drift. However, as research and development progresses, the long-term stability of these devices might improve; in this event, the gyroscopic properties of an AAA would similarly negligible drift, similar to strictly vertical atomic accelerometers \cite{niebauer1995new, bidel2020absolute, oon2022compact, lautier2014hybridizing, yankelev2019multiport}.

The size of the proposed AAA is suitable for navigation application of ships and submarines, however it may become an obstacle for smaller vehicles. The separation distances proposed in the long-term stability analysis, Figure~\ref{fig:DriftAnalysis}, of $l=1\;$m alludes to a sensor which requires precision and care in reducing the positional uncertainty and misalignment to a necessary level (Figure~\ref{fig:Resolution}), otherwise the error will begin to diverge once a factor surpasses some threshold value. Fortunately, these thresholds can be surpassed with modern technology \cite{peters2001high, oon2022vibration, d2017canceling, oon2022vibration}. It should be noted that the requisite separation of $l=1$m is due to the choice of using $N=4$ three-axis sensors; as per Eq.~\eqref{eq:omega_resolution}, one can reduce $l$ by increasing $N$. This is ultimately restricted by the size of the individual sensors, and although atomic sensors are quite large, significant progress has been made on devising compact hardware in the past few decades \cite{cheinet2006compact, sorrentino2010compact, oon2022compact} - it is not far-fetched to assume that progress will continue in size reduction.

Unlike most classical inertial sensors, atomic sensors operate within a very narrow dynamic range. For example, atomic accelerometers typically operate within a bandwidth of $10^{-4}-10^{-3}$ms$^{-2}$ \cite{geiger2020high,tennstedt2023atom}. This is a necessity so that the atoms properly interact with the laser pulses \cite{geiger2020high, narducci2022advances, stern2009light}; consequently there is a trade-off between resolution and dynamic range \cite{geiger2020high}. The limited bandwidth of an AAA is not an obstacle for some potential applications, such as stabilization \cite{algrain1993accelerometer, latt2011placement, lee2003increasing}, seismology \cite{niazi1986inferred, oliveira1989rotational, spudich1995transient, huang2003ground}, and perhaps even navigation of slower moving vehicles. However, for general navigational purposes, the AAA would be expanded to a hybrid system containing classical sensors.  In fact, atomic inertial sensors are often hybridized with classical sensors \cite{lautier2014hybridizing, cheiney2018navigation, geiger2011detecting, merlet2009operating, tennstedt2023atom} to improve sensitivity and reduce the amount of dead time between measurements. The bias drifts inherit to classical sensors are then filtered out through the means of statistical methods, such as a Kalman filter \cite{lautier2014hybridizing, cheiney2018navigation, huang2022atomic, tennstedt2023atom}. By incorporating sufficiently accurate classical sensors, one could devise an adaptive and predictive scheme to ensure the AAA is kept within a functional dynamic range \cite{tennstedt2021integration, tennstedt2023atom}.

In this manuscript, we have established the theoretical foundations of using an AAA to measure rotational velocity. This has many interesting applications, such as navigation \cite{tan2001design, zorn2002gps, buhmann2006gps, qin2006attitude, naseri2014improving}, stabilization \cite{algrain1993accelerometer, latt2011placement, lee2003increasing} and seismology \cite{niazi1986inferred, oliveira1989rotational, spudich1995transient, huang2003ground}. Nonetheless, it is crucial to test the validity with an experimental demonstration. The simplest method would consist of two three-axis atomic accelerometers, which could measure a single vector component of $\bm{\omega}$, which will suffice as a proof-of-principal demonstration.

\bibliography{main}

\section*{Acknowledgements}

This research is supported by the National Research Foundation, Singapore and A*STAR under its Quantum Engineering Programme (NRF2021-QEP2-03-P01 and NRF2021-QEP2-03-P06) and the DSO National Laboratories. The authors would also like to thank Fong En Oon, Kai Sheng Lee, Elizaveta Maksimova and Christoph Hufnagel for fruitful discussions.

%\section*{Author contributions statement}
%N. S. conducted the statistical analysis.  All authors reviewed the manuscript.

\section*{Data Availability}
The raw data used in Figures 2-4 can be downloaded from the following Github repository: https://github.com/GyroEmulator/Raw-Data.

\section*{Conflict of interest}
The authors have no conflicts to disclose.

\end{document}